\documentclass[graphics,floatfix, footinbib,tightenlines,nobibnotes, aps, prb, twocolumn]{revtex4-1}
\usepackage{amsmath,amssymb,wasysym}
\usepackage{graphicx}
\usepackage{dcolumn}
\usepackage{bm}
\usepackage{braket}
\usepackage{subcaption}
\usepackage{verbatim}
\usepackage{float}
\usepackage{bbold}
\usepackage{color}
\usepackage{xcolor}
\usepackage{relsize}
\usepackage{amsthm}
\usepackage{enumerate}
\usepackage{soul,xcolor}
\usepackage[T1]{fontenc}
\usepackage[colorlinks=true ,urlcolor=blue,urlbordercolor={0 1 1}]{hyperref}

\newcommand{\beq}{\begin{eqnarray}}
\newcommand{\eeq}{\end{eqnarray}}

\newcommand{\bsp}{\begin{split}}
\newcommand{\esp}{\end{split}}

\newcommand{\be}{\begin{equation}}
\newcommand{\ee}{\end{equation}}

\begin{document}

\setstcolor{red}

\title{Electrical detection of spin liquids in double moir\'e layers}
\author{
Ya-Hui Zhang}
\author{Ashvin Vishwanath}
\affiliation{Department of Physics, Harvard University, Cambridge, MA, USA
}

\date{\today}

\begin{abstract}
Although spin is a fundamental quantum number, 
measuring spin transport in traditional solid state systems is extremely challenging. This poses a major obstacle to detecting interesting quantum states including certain spin liquids. 
In this paper we propose a platform that not only allows for the electrical measurement of spin transport, but in which a variety of exotic quantum phases may be stabilized. Our proposal involves two moir\'e superlattices, built from  transition metal dichalcogenides (TMD) or  graphene, separated from one another by a thin insulating layer. The two  Coulomb coupled moir\'e layers, when suitably aligned, give rise to a layer  pseudospin degree of freedom. The  transport of pseudospin  can be accessed from purely electrical measurements of counter-flow  or Coulomb drag conductivity. Furthermore, these platforms naturally realize Hubbard models on the triangular lattice with $N = 4\, {\rm or}\, 8$ flavors. The flavor degeneracy motivates a large-N approximation from which we obtain the phase diagram of  Mott insulators at different electron fillings and correlation strengths.
In addition to conventional phases such as psuedospin superfluids and crystallized insulators, exotic phases including chiral spin liquids and a $U(1)$ spinon Fermi surface spin liquid are also found,  all of which will show  smoking gun electrical signatures in this setup.  
\end{abstract}

\pacs{Valid PACS appear here}
\maketitle


It is now well appreciated that spin plays an important role in strongly correlated systems. In addition to simple ferromagnetic or anti-ferromagnetic ordered phases, electronic spins can form non-ordered phases such as spin liquids\cite{anderson1973resonating,anderson1987resonating,kalmeyer1987equivalence,balents2010spin,savary2016quantum,knolle2019field}, which host fractional excitations.  Spin liquids can also exhibit non-trivial spin transport. For example, a spin liquid with a  spinon Fermi surface supports metallic spin conductivity while the chiral spin liquid displays a quantized fractional spin Hall conductivity, a spin analog of the electrical fractional quantum Hall effect (FQHE)\cite{kalmeyer1987equivalence,wen1989chiral}. Certain chiral superconductors also display a quantized spin quantum Hall effect  \cite{read2000paired,senthil1999spin}. Spin-charge separation has been proposed to occur in the metallic pseudogap phase of hole doped cuprates\cite{lee2006doping,senthil2003fractionalized,zhang2020pseudogap}. Thus a direct probe of spin transport could provide  important clues to cuprate physics and a way to detect a host of novel phases. Alas,  measuring spin transport in traditional solid state systems is unfeasible.  In ultracold atomic gases however, spin conductivity can be measured in the spinful Hubbard model since the role of spin is played by effective pseudospin  which allows for a greater degree of control\cite{nichols2019spin}. Even so, other issues notably with cooling to very low effective temperatures, crop up in the atomic platforms. Here, we propose to measure the transport of a pseudospin formed by the layer degree of freedom in an electronic material.

Our proposal is based on two recent experimental advances. First, in recently fabricated moir\'e systems, correlated insulators have been observed in several moir\'e systems based on graphene\cite{cao2018correlated,Wang2019Signatures,Wang2019Evidence,yankowitz2019tuning,Wang2019Signatures,chen2019tunable,lu2019superconductors,Cao2019Electric,Liu2019Spin,Shen2019Observation,polshyn2020nonvolatile,chen2020electrically} and transition metal dichalcogenides (TMD) heterobilayer\cite{tang2020simulation,regan2019optical,wang2019magic}. However, the  measurement of these moir\'e systems is limited to the charge transport and it is hard to probe the spin physics of the correlated insulators.   Second, we note a recent quantum Hall experiment \cite{LiuKim} in which two parallel sheets of graphene separated by an thin insulating hBN layer were found to realize novel interlayer correlated states which were established from their  Coulomb drag signatures.  Here we propose a  new class of moir\'e system which not only adds a new platform to explore  correlated physics, but also makes possible an {\em electrical} measurement of pseudo-spin transport, corresponding to the layer degree of freedom.

We will show that double moir\'e layer systems can simulate, to a good approximation, the $SU(N)$  Hubbard model on the triangular lattice with $N=4$ and $N=8$. Then we will focus on the Mott insulating regime and derive a spin model with ring exchange terms up to $O(\frac{t^4}{U^3})$ for general $N$. We map out the phase diagram for various filling $\nu$ while retaining a three-site ring exchange term $K$ which is pertinent to models with $N>2$. In  the Heisenberg limit ($K=0$), our results on the triangular lattice are  similar to those of a previous study on the square lattice\cite{hermele2009mott}.  The effect of the $K$ term was, however, not explored before and we find that reasonable values of $K$ can favor a chiral spin liquid (CSL) or a $U(1)$ spin liquid with spinon Fermi surface even the ground state is in a crystallized phase at $K=0$.  Thanks to the ability of measuring the pseudospin conductivity electrically,  both CSL and spinon Fermi surface states can be easily detected with clear signatures in our setting.

\begin{figure}
\centering
\includegraphics[width=0.45\textwidth]{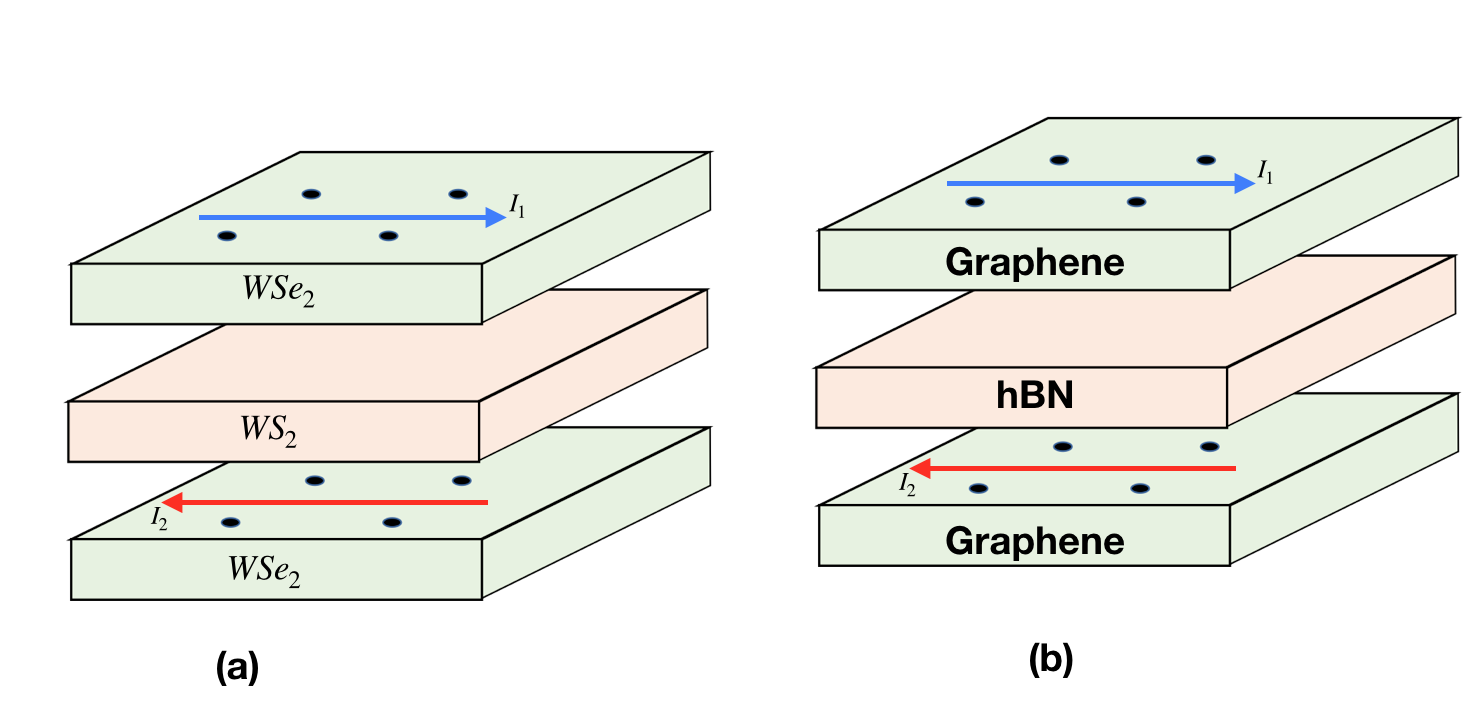}
\caption{ Double moir\'e layers systems formed by (a) WSe$_2$-WS$_2$-WSe$_2$ and (b) graphene-hBN graphen.  We assume alignment i.e. the twist angle is $0$ between the top(bottom) layer and the middle layer. The middle layer has two roles: (1) provides a moir\'e superlattice for the top and the bottom layer. (2) provides an insulating barrier to suppress the tunneling between the top and the bottom layer.  In this setup, one can apply electric field $E_a$ in layer $a$ and measure the induced current $I_b$ in layer $b$. }
\label{fig:Moiredouble layer}
\end{figure}

 {\bf Setup:} As a concrete example, we consider WSe$_2$-WS$_2$-WSe$_2$ heterostructure as shown in Fig.~\ref{fig:Moiredouble layer}. Because of the lattice constant mismatch, WS$_2$ provides a triangular moir\'e superlattice for the two WSe$_2$ layers. It has already been demonstrated experimentally\cite{tang2020simulation,regan2019optical} that the  moir\'e narrow band in the case of the single WSe$_2$ layer adjacent to WS$_2$ can simulate a spinful Hubbard model and realize a Mott insulator with a large charge gap $\Delta_c \approx 10 $ meV at half filling. In our proposed system, there is an additional pseudospin formed by the layer degree of freedom, which, combined with the spin, leads to an approximate $SU(4)$ Hubbard model for the narrow band. Because of the fairly good inversion symmetry of TMD at $K$ and $K'$ points, there is no flux in the hopping and no Berry curvature in  momentum space, which generically occurs for moir\'e bands from graphene\cite{zhang2019bridging}.  Similarly, we can also consider graphene-hBN-graphene heterostructure, which has 8 flavors formed by spin, valley and layer.  For the graphene layer, we can use ABC trilayer graphene (TG).  For  TG-hBN-TG,  the displacement fields $D_1$ and $D_2$ within each graphene can be used to tune the bandwidth and even the band topology of the moir\'e band\cite{zhang2019nearly}. In addition, we can tune the total density $N=N_1+N_2$ and layer spin $\rho_z=N_1-N_2$ separately. In total there are four variables $N,\rho_z, D_1, D_2$ to explore and a rich phase diagram is expected.

{\bf Low energy model:} We consider systems as shown in Fig.~\ref{fig:Moiredouble layer}. In the ideal case we want the two triangular moir\'e superlattices to be perfectly aligned. In practice, it is possible to align the angle, but the relative shift $R_{\rm shift}$ of the two lattices is hard to control externally.  However, as we argue later, a small $R_{\rm shift}$ does not change the physics qualitatively.  A large $R_{shift}$ will lead to a different lattice model, which may also be interesting.  In this paper we will restrict to small $R_{\rm shift}$ for simplicity. In this case we can always use the same index $i$ to label the sites of lattices in both layers. We will view the layer index $a$ as internal degree of freedom, which combines with the spin-valley index $\alpha$ to form the spin index of an effective Hubbard model\cite{wu2018hubbard,zhang2019bridging,zhang2020spin} at low energy:

\begin{equation}
H=- \sum_{\langle ij \rangle} t_{a\alpha} c^\dagger_{i;a\alpha}c_{i;a\tilde \alpha}+\frac{U}{2}\sum_a  n_{i}^2- \delta \sum_i n_{i;1}n_{i;2}+...
\end{equation}
where $a=1,2$ labels the two layers.

We include interaction terms beyond on-site coupling in the $...$  and  $n_i=\sum_{a \alpha}c^\dagger_{i;a\alpha}c_{i;a\alpha}$.  If $d=0$ and $|R_{\rm shift}|=0$, the inter-layer on-site repulsion is the same as the intra-layer on-site repulsion and we can organize them together in a single on-site Hubbard $U$. In reality there is small $d$ and $\vec R_{shift}$, thus we need to add a small $\delta$ term. We will also assume that the two layer have the same hopping. For TMD double layer, $\alpha=\uparrow, \downarrow$ and in total we have an approximate $SU(4)$ Hubbard model. For graphene-hBN-graphene, $\alpha=K\uparrow, K\downarrow, K'\uparrow, K'\downarrow$ is the spin-valley index. In this case the hopping term has valley-contrasting flux. For simplicity, in this paper we will ignore the valley-contrasting flux and consider a $SU(8)$ model formed by layer-spin-valley. We label the Pauli matrices for layer, spin, valley as $\rho_a$, $\sigma_a$, $\tau_a$.  In the remaining part of the paper we will focus on the following $SU(N)$ Hubbard model with $N=4,8$ on triangular lattice:
\begin{equation}
 	H=-t \sum_{\alpha,\langle ij\rangle} c^\dagger_{i;\alpha} c_{j;\alpha}+\frac{U}{2} \sum_i n_i^2 +\delta \sum_i\rho_{i;z} \rho_{i;z}
 \end{equation} 

 The $\delta$ term introduces easy-plane anisotropy which favors the super-spin to point along the in-plane direction in  layer space.  For TMD double layer, it breaks the $SU(4)$ symmetry down to $SU(2)_1 \times SU(2)_2$, where $SU(2)_a$ is the spin rotation for layer $a$.

{\bf Electrical measurement of  layer psuedospin transport:} The double moir\'e layer system makes it possible to measure the pseudospin transport corresponding to the layer $\rho_z$.  Because both $N$ and $\rho_z$ are conserved quantities, we can define current $\vec J_c$ and $\vec J_s$ for them. In the language of the Hubbard model we are trying to simulate, $\vec J_c$ is the charge current and $\vec J_s$ is a spin current. 

In the experiment, because the two layers are separated by an insulating layer, it is possible to apply electric field $E_a$ in one layer and measure the current $j_b$ in another layer. In this way, we can get $\sigma^{ab}_{xx}=\frac{j^a_x}{E^b_x}$ and $\sigma^{ab}_{xy}=\frac{j^a_x}{E^b_y}$.  With redefinition $\vec J_{c,s}=\vec{J_1}\pm \vec{J_2}$ and $\vec E_{c,s}=\frac{\vec{E_1}\pm \vec{E_2}}{2}$, we can also apply charge (pseudospin) electric field $\vec E_c$ ($\vec E_s$) and measure the charge (pseudospin) current $\vec J_c$ ($\vec J_s$).  Therefore it is possible to measure $\sigma^{ss}_{xx}, \sigma^{cc}_{xx}, \sigma^{sc}_{xx}$ and $\sigma^{ss}_{xy}, \sigma^{cc}_{xy}, \sigma^{sc}_{xy}$. Now  $\sigma^s_{xx}=\sigma^{ss}_{xx}$ and $\sigma^s_{xy}=\sigma^{ss}_{xy}$ are the pseudospin conductivity and the pseudospin Hall conductivity. 

Even in the Fermi liquid, $\sigma^s$ can be different from $\sigma^c$, thou the effect is expected to be small. 
In the strongly correlated regime, spin and charge may separate and $\sigma^s$ can be even more significantly different from charge conductivity $\sigma^c$.  The best established regime for such physics is the Mott insulator at integer filling, where the charge is frozen by the large $U$. In a Mott insulator, $\sigma^c_{xx}=\sigma^c_{xy}=0$, but the spin can still be active  and have non-trivial spin transport.


In contrast to the quantum Hall bilayers, the Hubbard model in our case provide antiferromagnetic spin coupling in the Mott insulator regime. On triangular lattice, more exotic spin liquid phases may emerge and show non-trivial spin transport. In this paper we will show that two spin liquids can be favored and "smoking gun" evidence of them can be easily obtained from measurement of $\sigma^s_{xx}$ and $\sigma^s_{xy}$.

{\bf Large N calculation in Abrikosov fermion mean field theory} At integer filling $\nu_T$, the system is in a Mott insulator in the $U>>t$ limit because the number of electron is frozen at each site and the low energy physics is described by a spin model obtained from $t/U$ expansion. The results up to the $\frac{t^4}{U^3}$ can be found in the supplementary for general $N>2$. Here we keep only the $O(\frac{t^3}{U^2})$ term for the large N mean field calculation. Let us point out a special feature that appears for $N\ge 3$  which  is absent in the $SU(2)$ case - i.e. a new three-site ring exchange term $K$ even at zero magnetic field. For the SU(2) case with $S=1/2$ per site, this term  (at zero external magnetic field) just modifies the nearest neighbor Heisenberg coupling. This follows since two of the three spin projections in $N=2$ must be identical. However, for SU(N) $N>2$, a novel three spin ring interaction $K$ is generated, which we explicitly include. 

 We introduce fermionic spinon $f_{i;\alpha}$ at each site with $\alpha=1,2,...,N$ to denote the spin degree of freedom. The constraint is
\begin{equation}
	\sum_\alpha f^\dagger_{i;\alpha}f_{i;\alpha}=\nu_T
\end{equation}

Equivalently the density of each flavor is $\nu=\frac{\nu_T}{N}$.  For simplicity we only keep the spin model to the third order of $t/U$:
\begin{align}
H_S&=-J\sum_{\langle ij \rangle}f^\dagger_{i;\alpha}f_{j;\alpha} f^\dagger_{j;\beta} f_{i;\beta}\notag\\
&-K \sum_{i,j,k\in \bigtriangleup}\left(f^\dagger_{i;\alpha}f_{k;\alpha}f^\dagger_{k;\gamma}f_{j;\gamma}f^\dagger_{j;\beta}f_{i;\beta} e^{i \Phi_3}+h.c.\right)
\label{eq:spin_model}
\end{align}
with $J=2\frac{t^2}{U}$ and $K=6\frac{t^3}{U^2}$. $\Phi_3$ is the external magnetic flux through a triangle.

We can use a mean field theory to get the phase diagram with $K$ and $\nu$. The mean field theory ignores fluctuations and is well known to fail for $SU(2)$ case. However, in the $N\rightarrow \infty$ limit the fluctuation around the saddle point is suppressed by the integration of fermions and mean field theory becomes more accurate.  Therefore, a mean field calculation may be reliable for our $N=8$ case and may also provide useful insights for the $N=4$ case. 

To have a controlled large $N$ calculation, we need to rescale $\tilde K=KN^2$ and $\tilde J=JN$. We will fix the filling $\nu=\nu_T/N$, $\tilde K$ and $\tilde J$ when taking $N$ to infinite.

We can decouple the spin model to have a mean field ansatz:
\begin{equation}
	H_M=-\sum_{\langle ij \rangle}\chi_{ij} f^\dagger_{i\alpha}f_{j\alpha}+h.c.-\sum_i \mu_i f^\dagger_{i;\alpha}f_{i;\alpha}
	\label{eq:mean_field}
\end{equation}
where $\mu_i$ is introduced to satisfy the constraint: $\frac{1}{N}\sum_\alpha \langle f^\dagger_{i;\alpha}f_{i;\alpha}\rangle=\nu$.

Mean field ansatz $\chi_{ij}$ can be obtained by Feynman's variational principle\cite{brinckmann2001renormalized}. Basically the free energy $\beta F\leq \Phi[\chi_{ij}]$, here
\begin{equation}
\Phi=\langle S -\tilde S \rangle- \log \tilde Z	
\end{equation}
where $S$ is the action of the full Hamiltonian and $\tilde S$ is the action corresponding to the mean field ansatz in Eq.~\ref{eq:mean_field}. $\tilde Z=\int D[f] e^{-\beta \tilde S}$ is the partition function of the mean field theory.  

We can obtain mean field ansatz $\chi_{ij}$ by minimizing $\Phi[\chi_{ij}]$, which leads to self consistent equations:

\begin{equation}
	\chi_{ij}=\tilde J\langle T_{ji} \rangle+\tilde K \sum_{k,i,j \in \bigtriangleup}e^{-i \Phi_3}\langle T_{jk} \rangle\langle T_{ki} \rangle 
	\label{eq:self_consistent_equations}
\end{equation}
where $T_{ij}=\frac{1}{N}\sum_\alpha f^\dagger_{i\alpha} f_{j\alpha}$.

At $T=0$, variational energy is:
\begin{equation}
	\frac{E_M}{N}=-\tilde J\sum_{\langle ij \rangle}\langle \hat T_{ij} \rangle \langle T_{ji} \rangle-\tilde K\sum_{i,j,k\in \bigtriangleup}(e^{-i\Phi_3} \langle T_{ik}\rangle \langle T_{kj} \rangle  \langle T_{ji}\rangle +h.c.)
	\label{eq:mean_field_energy}
\end{equation}

In the above we have substituted $J=\frac{\tilde J}{N}$ and $K=\frac{\tilde K}{N^2}$ to Eq.~\ref{eq:spin_model}.

\begin{figure}
\centering
\includegraphics[width=0.45\textwidth]{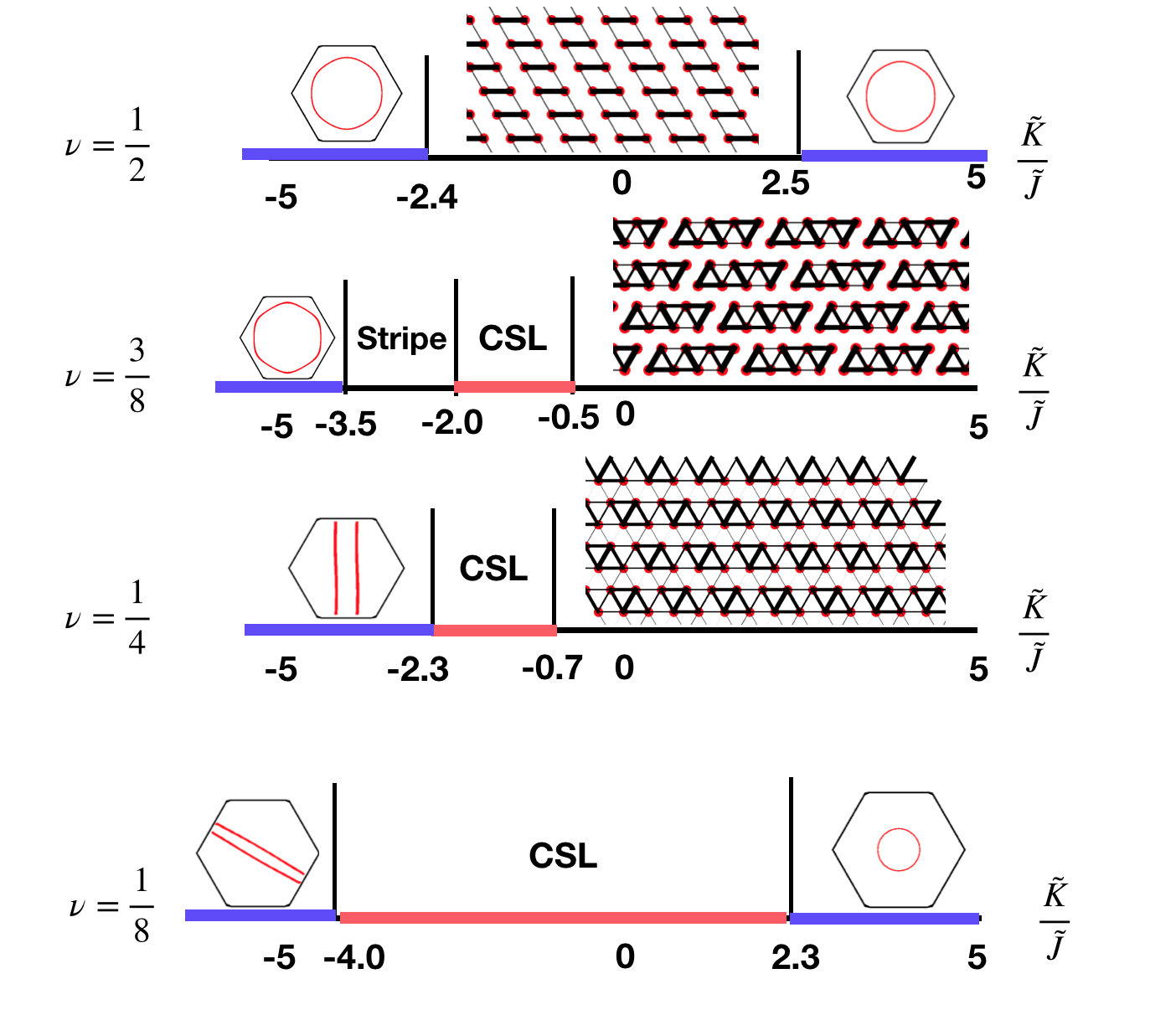}
\caption{Phase diagram from the large $N$ calculation.  CSL denotes chiral spin liquid. The pattern of $|\chi_{ij}|$ for various crystal phases are also shown. In the large $\tilde K$ regime, we also find $U(1)$ spin liquids with spinon Fermi surfaces as shown.}
\label{fig:large_N_phase_diagram}
\end{figure}

We show details of the calculation in the supplementary. We can always make a particle hole transformation $\nu\rightarrow 1-\nu, t \rightarrow -t$. Hence we only show results for $\nu \leq \frac{1}{2}$. A phase diagram is shown in Fig.~\ref{fig:large_N_phase_diagram} with $\nu$ and $\frac{\tilde J}{\tilde K}$. At the Heisenberg limit with $\tilde K=0$, we find the ground state is crystallized for $\nu=\frac{1}{4},\frac{1}{2},\frac{3}{8}$, but is a chiral spin liquid for $\nu=\frac{1}{8}$.  This result is similar to the large-N calculation on square lattice\cite{hermele2009mott}.  The crystal phase is a valence bond solid (VBS) for $\nu=\frac{1}{2}$ and a Plaquette order for $\nu=\frac{1}{4}$ (see the supplementary).   In our calculation, we find the ring exchange term $\tilde K$ can destabilize the crystal phase and favor the chiral spin liquid or $U(1)$ spin liquid with spinon Fermi surfaces. For example, at $\nu=\frac{1}{4}$, $\frac{\tilde K}{\tilde J}=N\frac{K}{J}<-0.7$ is enough to favor a chiral spin liquid over the plaquette order.  As $\frac{K}{J} \approx 3 \frac{t}{U}$, we need a critical value of $\frac{t}{U}<-\frac{0.23}{N}=-0.057$ for $N=4$, which is still in the strong Mott regime. The chiral spin liquid is only found for negative $t$. For a Hubbard model with positive $t$, we should search for the CSL at the $\nu=\frac{3}{4}$ filling.  We also find a $U(1)$ spin liquid with spinon Fermi surfaces in the large $|\tilde K|$ regime. They are always translation invariant, but some ansatz breaks rotation symmetry and has hopping only along one direction, resulting a  decoupled chain phase.

The CSL phase has uniform amplitude, but the phase of the hopping breaks translation symmetry. For example, for $\nu=\frac{1}{8}$ ($\nu=\frac{1}{4}$), the ansatz has one flux in the enlarged $4 \times 2$  ($2 \times 2$) unit cell and therefore each flavor occupies a Chern band with $C=1$.  For $\nu=\frac{3}{8}$, each flavor occupies $3$ bands resulting from a $4 \times 2$ unit cell. The total Chern number of the occupied three bands is $C=1$.

{\bf Electrical signatures for spin physics}

In this section we discuss how we detect various spin phases found in the large N calculation for the Mott insulator.We will discuss the psudospin transport signatures of a variety of both exotic and conventional phases ranging from psuedospin chiral liquids with quantized psuedospin Hall transport, non-Fermi liquid metallic psuedospin transport in spinon Fermi surface states, as well as psuedospin superfluid phases. In all cases the electric transport remains insulating.

{\bf  Crystallized order and exciton condensation driven by displacement field}

At filling $\nu=\frac{1}{4}$, $\nu=\frac{3}{8}$ and $\nu=\frac{1}{2}$, we find the ground state in the strong Mott limit is a crystallized phase.  For $\nu=\frac{1}{2}$, the gap for the VBS state is $\Delta=\tilde J=2J$ for $N=2$. For $\nu=\frac{1}{4}$, the gap for the plaquette order  is $\Delta=0.4 \tilde J=1.6 J$ for $N=4$. Therefore $\sigma_s$ measured from counterflow transport should be zero at zero temperature. Next we show that inter-layer coherence can be driven by displacement field $D$.  On top of the crystallized phase, there are gapped excitations generated by $S^\alpha_\beta=f^\dagger_\alpha f_\beta$. Let us focus on the $SU(4)$ model formed by double TMD layer.  These gapped excitations can be grouped to $ \psi^\dagger \rho_a \sigma_b \psi \ket{G}$ with $a,b=0,1,2,3$.  Next we add a term:

\begin{equation}
	H=-D \sum_i \psi^\dagger_i \rho_z \psi_i
\end{equation}

Then the gapped magnon corresponding to $\rho^{+}$ and $\rho^{+} \vec{\sigma}$ will have energy: $E(D)=\Delta-2D$.  The magnons corresponding to $\rho^-,\rho^-\vec{\sigma}$ have the energy $E(D)=\Delta+2D$. The magnons generated by $\rho_{0,z}\sigma_a$ has constant energy with $D$. As a result, when $D>\frac{\Delta}{2}$, the magnons generated by $\rho_{+}$ and $\rho_{+} \vec{\sigma}$ will condense.  The magnons from $\rho^+$ and $\rho^+ \vec{\sigma}$ form a $SO(4)$ order parameter. The final state will be selected from this $SO(4)$ manifold by some small parameters. But anyway the $U(1)$ symmetry generated by $\rho_z$ is broken and we have inter-layer coherence. As a result, the state will behave as a superfluid in the counter-flow transport. This is the same symmetry breaking order as the exciton condensation phase in the quantum Hall bilayer. But in our setting up, the BEC of magnons can happen at zero magnetic field and is driven by a displacement field.

{\bf Chiral Spin Liquid: counter-flow Hall conductivity}

For $SU(N)$ model, the CSL state has $N$ flavors, each is in a Chern insulator with $C=1$. 

We introduce an external gauge field $A_s=\frac{A_1-A_2}{2}$. The action for the CSL phase is

\begin{align}
 	L&=-\sum_{I=1}^{N}\frac{1}{4\pi}\alpha_I d \alpha_I+\frac{1}{2\pi} \sum_{I=1}^{N} a d \alpha_I\notag\\
 	&+\frac{1}{2\pi} (\sum_{I=1}^{\frac{N}{2}}A_s d \alpha_I -\sum_{I=\frac{N}{2}+1}^N A_s d \alpha_I)
 \end{align} 
 where $\alpha_I$ is introduced to describe the Chern insulator phase for the flavor $I$. $a$ is the $U(1)$ gauge field from the Abrikosov fermion parton construction.  $\alpha_I$ with $I\leq \frac{N}{2}$ and $I>\frac{N}{2}$ correspond to the top and the bottom layer respectively.

 The above action has $U(1)_N$ topological order  with the underlying anyon excitations carrying self statistics $\theta=\pi-\frac{\pi}{N}$. These transform in the fundamental representation of the SU(N) global symmetry and give rise to  
 a spin Hall conductivity $\sigma^s_{xy}=N \frac{e^2}{h}$ corresponding to the response $\frac{N}{4\pi} A_s d A_s$.   A current $I^s_x=2I^1_x=-2 I^2_x=I$ will induce a voltage $V^1_y=-V^2_y=V$ with spin Hall resistivity $R^s_{xy}=\frac{V}{I}=\frac{1}{N} \frac{h}{e^2}$.

The energy scale below which CSL behavior is manifest is related to the gap in our theory which is of order $J$. More precisely at $\nu=\frac{1}{8}$ when CSL is realized even when  $\tilde K=0$, the gap we obtain is $\Delta=0.2 \tilde J=1.6 J$ for $N=8$. For $\nu=\frac{1}{4}$ at $\frac{\tilde K}{\tilde J}=-1$, the CSL gap is $\Delta=0.4 \tilde J=1.6 J$ for $N=4$ while for $\nu=\frac{3}{8}$ at $\frac{\tilde K}{\tilde J}=-0.5$, the CSL gap is $\Delta=0.6 \tilde J=4.8 J$ for $N=8$, which appears with a relatively large prefactor.

{\bf Spinon Fermi surface: metallic pseudospin conductivity and quantum oscillation}

As we show, increasing $t/U$ can drives the Mott insulator into a $U(1)$ spin liquid with spinon Fermi surfaces. The low energy physics of this phase is described by spinon Fermi surfaces with $N$ flavors coupled to a $U(1)$ gauge field. In $2+1$ dimension, such a theory is strongly coupled and even the $\frac{1}{N}$ expansion fails. It was believed that the theory flows to a non-Fermi-liquid fixed point in low temperature, although a well-established theory is still absent despite lots of efforts\cite{lee2018recent}.  Here we do not attempt to solve this long time problem. Instead, we show that the counter-flow measurement of  the pseudospin conductivity can provide important information to test predictions of the non-Fermi-liquid behaviors.  Through counter-flow measurement we can get pseudospin resistivity $R(T)$ at any temperature and one can extract the exponent $\alpha$ from fitting $R(T)\propto T^\alpha$ at low temperature.  Due to the strong gauge fluctuation, it is expected that $\alpha$ deviates from the Fermi liquid value (which is $2$).  An experimental measurement of the exponent $\alpha$ is definitely valuable for better understanding of this exotic phase. Besides, the spinon Fermi surface state should also have a spin analog of Shubnikov-de Haas quantum oscillation under external magnetic field $B$ (see the supplementary).


{\bf Enhanced Coulomb drag in metallic regimes}

Lastly we turn our attention away from the Mott insulator to the metallic regime. Now both $\sigma^s$ and $\sigma^c$ are non-zero.   It can be easily show that  $\rho_c-\rho_s$ is actually the Coulomb drag resistivity. If we apply current $I$ in the second layer, the induced resistivity $V$ in layer 1 is determined by $\rho_{D}=\frac{V}{I}=\rho_c-\rho_s$.  Strictly speaking, $\rho_c$ can be different from $\rho_s$ even in the Fermi liquid because there is no symmetry relating the charge current to the pseudospin current. However, the Drag resistivity $\rho_D$  in Fermi liquid theory needs electron-hole asymmetry and is generically a small effect\cite{narozhny2016coulomb}.  Here we propose a much more significant Drag effect due to the spin-charge separation or symmetry breaking in metals close to a Mott insulator.

First, we can look at the Mott transition tuned by $\frac{t}{U}$, which can be gate tunable in ABC trilayer graphene\cite{zhang2019bridging}. One possible scenario of a continuous Mott transition is described by the slave boson theory\cite{senthil2008theory} $c_{i;\sigma}=b_i f_{i;\sigma}$. In this theory we have $\rho_c=\rho_b+\rho_f$, while $\rho_s\approx \rho_f$, where $\rho_b$ and $\rho_f$ are the "charge" resistivity for the slave boson $b$ and the spinon $f$.  Across the transition,  the slave boson $b$ goes through a superfluid-Mott transition and as a result $\rho_b$ jumps from $0$ to infinite, while $\rho_f$ is smooth.  Exactly at the critical point, $b$ has a universal resistivity $R_U$ at order of $\frac{h}{e^2}$. Therefore $\rho_c=R_U+\rho_f\approx \rho_s+R_U$ and we have a large drag resistivity $\rho_D\approx R_U$ at order of $\frac{h}{e^2}$.

Another place to search for a possible large drag effect is in the underdoped regime away from the Mott insulator. Here there may be an exotic metallic state with spin-charge separation, such as a fractional Fermi liquid. In this kind of phase there is a spin liquid part in addition to the Fermi liquid, which may contribute an additional resistivity or Hall resitivity to $\rho_s$, leading to drag or Hall drag effect.  Another possibility is a magnetic ordered metal with inter-layer coherence. In this case we have $\rho_s=0$ and as a result $\rho_D=\rho_c$.

While there is no  well established theory for the metal insulator transition and for the doped Mott insulator, here we content ourselves with 
pointing out that the drag measurement in the double moir\'e layer setting can provide  valuable information which can help unveil mysteries in the  metallic regime close to the Mott insulator.

{\bf Conclusion}
In conclusion, we propose a new system formed by two moir\'e layers coupled together with Coulomb interaction.  This set up enables electrical measurement of the transport corresponding to the layer pseudo-spin. The simple double moir\'e layers formed by TMD or ABC trilayer graphene can simulate approximate $SU(4)$ or $SU(8)$ Hubbard model. Based on large $N$ calculation, we predict the existence of crystallized phase, chiral spin liquid and $U(1)$ spin liquid with spinon Fermi surfaces in the Mott insulators. All of these three phases can be detected by the counter-flow measurements we proposed.  The ability of measuring pseudo-spin conductivity may also shed light on the possible topological superconductor or unconventional metals emerging from doped  Mott insulators. A target for future work is to create moir\'e platforms that simulate strongly correlated systems like the square lattice Hubbard model, along with the ability to probe spin conductivity. 

{\bf Acknowledgements:} We thank Philip Kim and Dapeng Ding for discussions. This work was supported by the Simons Collaboration on Ultra-Quantum Matter, which is a grant from the Simons Foundation (651440, AV).

\bibliographystyle{apsrev4-1}
\bibliography{double_moire}

\onecolumngrid

\appendix

\section{Spin model from $t/U$ expansion}
At integer total filling $\nu_T$,  we can perform the standard $t/U$ expansion to get an effective spin model for the Mott insulator in the large U limit.   The idea is to divide the kinetic term in terms of the change of interaction energy (for spin 1/2 case, it is just the change of double occupancy number)\cite{macdonald1988t}.  We will illustrate the calculation for the $SU(4)$ case. However, up to the $\frac{t^4}{U^3}$ order, the result is the same for any $N\geq 4$.

We consider the Hamiltonian:
\begin{equation}
	H=T+H_U
\end{equation}

For $SU(4)$,  we can write the hopping term as:
\begin{equation}
	T=T_0+T_1+T_{-1}+T_2+T_{-2}+T_3+T_{-3}
\end{equation}

where $T_m$ increases the interaction energy by $mU$. Or in another language:
\begin{equation}
	[H_U,T_m]=m  U T_m
\end{equation}
where $H_U=\frac{U}{2}\sum_i n_i^2$ is the Hubbard interaction.

We label $T_{pq}=-t \sum_{ij;a}  c^\dagger_{i;a} c_{j;a} P_{i;p} P_{j;q}$, where $P_{i,p}$ is a projector which constraints $n_i=p$. Then

\begin{align}
T_0&=T_{01}+T_{12}+T_{23}+T_{34} \notag\\
T_1&=T_{11}+T_{22}+T_{33}\notag\\
T_{-1}&=T_{02}+T_{13}+T_{24}\notag\\
T_{2}&=T_{21}+T_{32}\notag\\
T_{-2}&=T_{03}+T_{14}\notag\\
T_{3}&=T_{31}\notag\\
T_{-3}&=T_{04}
\end{align}

It is easy to verify that $T_m^\dagger=T_{-m}$. We want to keep only $T_0$ and eliminate $T_m$ with $m\neq 0$ by a unitary transformation:

\begin{equation}
	H'=e^{i S}H e^{-i S}=H+[iS,H]+\frac{1}{2!}[iS,[iS,H]]+...
\end{equation}

We can perform the transformation systematically by order of $1/U$: $iS=iS^{(1)}+iS^{(2)}+iS^{(3)}+...$, where $iS^{(k)}$ is at order $O(\frac{1}{U^k})$.
We choose
\begin{equation}
	iS^{(1)}=\frac{1}{U}(T_1-T_{-1}+\frac{1}{2} T_2 -\frac{1}{2} T_{-2}+\frac{1}{3} T_3-\frac{1}{3} T_{-3})
\end{equation}

Because
\begin{equation}
	[iS^{(1)},H]=-(T_1+T_{-1}+T_2+T_{-2}+T_3+T_{-3})+\frac{1}{U}[T_1-T_{-1}+\frac{1}{2} T_2 -\frac{1}{2} T_{-2}+\frac{1}{3} T_3-\frac{1}{3} T_{-3}, T_0+T_1+T_{-1}+T_2+T_{-2}+T_3+T_{-3}]
\end{equation}

We get an effective Hamiltonian at order $O(\frac{t^2}{U})$:
\begin{align}
	H^{(2)}&=T_0+H_U+\frac{1}{U}[T_1-T_{-1}+\frac{1}{2} T_2 -\frac{1}{2} T_{-2}+\frac{1}{3} T_3-\frac{1}{3} T_{-3}, T_0+T_1+T_{-1}+T_2+T_{-2}+T_3+T_{-3}]\notag\\
	&-\frac{1}{2U}[T_1-T_{-1}+\frac{1}{2}T_2-\frac{1}{2}T_{-2}+\frac{1}{3}T_3-\frac{1}{3}T_{-3},T_1+T_{-1}+T_2+T_{-2}+T_3+T_{-3}]+O(\frac{1}{U^2})\notag\\
	&=T_0+H_U-\frac{1}{U}([T_{-1},T_1]+\frac{1}{2}[T_{-2},T_2]+\frac{1}{3}[T_{-3},T_3])+...
\end{align}

We can organize these terms by introducing the notation $T^{k}(m_k,...,m_1)=T_{m_k}T_{m_{k-1}}...T_{m_1}$ and $M(m_1,...,m_k)=\sum_{i=1}^k m_i$ associated with the sequence. In the above we only keep the terms with $M=0$.

We always have
\begin{equation}
	[H_U,T^{k}(m_k,...,m_1)]=M(m_1,...,m_k)U T^{k}(m_k,...,m_1)
\end{equation}

At second order, the effective spin model is
\begin{equation}
	H^{(2)}=-\frac{1}{U} T_{-1}T_1
\end{equation}
where we only keep terms which are not zero at the integer filling $\nu_T$. For doped case, more terms need to be kept.

At third order, We need to remove the other terms $T^{2}(m_2,m_1)$ with $M\neq 0$. This can be done by simply introducing
\begin{equation}
	iS^{(2)}=\frac{1}{U^2} \sum_{M(m_2,m_1)\neq 0}  C(m_2,m_1) \frac{1}{M(m_2,m_1)} T^{(2)}(m_2,m_1)
\end{equation}
which givs Hamiltonian $H^{(3)}$ from transformation generated by $iS=iS^{(1)}+iS^{(2)}$.  Here $C(m_2,m_1)$ is the coefficient of $T^{(2)}(m_2,m_1)$ in $H^{(2)}$.

Then at the fourth order, we need to remove the terms $T^{3}(m_3,m_2,m_1)$ with $M\neq 0$ generated in $H^{(3)}$.  Similarly we need to use 
\begin{equation}
	iS^{(3)}=\frac{1}{U^3} \sum  C(m_3,m_2,m_1) \frac{1}{M(m_3, m_2,m_1)} T^{(3)}(m_2,m_1)
\end{equation}

Finally, the unitary transformation generated by $iS=iS^{(1)}+iS^{(2)}+iS^{(3)}$ produces the effective Hamiltonian $H^{(4)}$ at the order of $O(\frac{t^4}{U^3})$. We numerically collect all of the terms and keep the only terms which are non-zero in the subspace at integer filling $\nu_T$, which should satisfy the necessary condition:
\begin{align}
&m_1=1\notag\\
&m_k=-1\notag\\
&\sum_{i=1}^n m_i \geq 0 
\end{align}
for any $n$.

The final result at the fourth order is:

\begin{align}
H^{(4)}&=-\frac{1}{U} T^{(2)}(-1,1)+\frac{1}{U^2}T^{(3)}(-1,0,1)+\frac{1}{U^3}(-T^{(4)}(-1,0,0,1)\notag\\
&-\frac{1}{2}T^{(4)}(-1,-1,1,1)+T^{(4)}(-1,1,-1,1)-\frac{1}{3}T^{(4)}(-1,-2,2,1))
\label{eq:spin_model_T_form}
\end{align}

The above expression is actually the same as the $SU(2)$ case. Actually the expression holds for any $SU(N)$.  Next we need to rewrite it using only spin operators. 

For $SU(N)$, we have generators $t^a$ with $a=1,...,N^2-1$ with normalization rule $Tr(t^a t^b) =4 \delta^{ab}$. We can then expand $T^{k}(m_k,...,m_1)$ with $t^{a_1}_1 t^{a_2}_2 ...t^{a_k}_k$. The coefficient is
\begin{equation}
	\frac{1}{N^k} Tr \big( t^{a_1}_1 t^{a_2}_2 ...t^{a_k}_kT^{k}(m_k,...,m_1)\big)
\end{equation}

For our purpose of doing large N calculation using Abrikosov fermion $f_\alpha$, it is more convenient to write the spin model directly with $f_\alpha$. Here for simplicity we only keep terms up to $O(\frac{t^3}{U^2})$, the spin model is

\begin{equation}
H=-\frac{2t^2}{U}\sum_{\langle ij \rangle}f^\dagger_{i;\alpha}f_{j;\alpha} f^\dagger_{j;\beta} f_{i;\beta}-6 \frac{t^3}{U^2} \sum_{i,j,k\in \bigtriangleup}\left(f^\dagger_{i;\alpha}f_{k;\alpha}f^\dagger_{k;\gamma}f_{j;\gamma}f^\dagger_{j;\beta}f_{i;\beta} e^{i \Phi_3}+h.c.\right)
\end{equation}
where we have assumed Einstein summation convention for $\alpha,\beta,\gamma=1,2,...,N$. $\Phi_3$ is the magnetic flux for each triangle.

The above expression works for any inter filling of $SU(N)$ model with $N \geq 3$.  One needs to be careful about the $SU(2)$ case. For $SU(2)$, the real part of the three-site ring exchange term can actually be decomposed to nearest-neighbor coupling. As a result, the only ring-exchange term is $-6 \frac{t^3}{U^2} i \sin(\Phi_3) \sum_{i,j,k\in \bigtriangleup}\left(f^\dagger_{i;\alpha}f_{k;\alpha}f^\dagger_{k;\gamma}f_{j;\gamma}f^\dagger_{j;\beta}f_{i;\beta}-h.c.\right)$, which vanishes without external magnetic flux.  For $N \geq 3$, we have the three-site ring exchange term even without external flux.

For the fundamental representation at $\nu_T=1$, we can also rewrite the spin model with ring-exchange terms, which are permutation of spin configuration.  We define $P_{ij}=\ket{\alpha \beta}\bra{\beta \alpha}$, $P_{ijk}=\ket{\alpha \beta \gamma}\bra{\beta \gamma \alpha}$ and $P_{ijkl}=\ket{\alpha \beta \gamma \delta}\bra{\beta \gamma \delta \alpha}$.  Here $\ket{\alpha_1 \alpha_2 \alpha_3 ... \alpha_k}$ labels one spin basis for the sites $i_1,i_2,i_3,...,i_k$.

We can rewrite the spin model in terms of ring-exchange terms:

\begin{align}
	H_S&= J \sum_{ij} P_{ij} +J' \sum_{\langle \langle ij \rangle \rangle} P_{ij} ++J'' \sum_{\langle \langle \langle ij  \rangle \rangle \rangle} P_{ij}\notag\\
	&+ \sum_{i,j,k \in \bigtriangleup} (K_3 P_{ijk}+h.c.) +\sum_{i,j,k \in \bigtriangleup'} (K_3' P_{ijk}+h.c.) \notag\\
	&+K_4 \sum_{i,j,k,l \in \square}(e^{i\Phi_4}P_{ijkl}+h.c.)
	\label{eq:spin_model_fourth_order_ring_exchange}
\end{align}
with 
\begin{equation}
	J=2 \frac{t^2}{U}+12\frac{t^3}{U^2}+60 \frac{t^4}{U^3}
\end{equation}
where we assume hopping is $-t c_i^\dagger c_j$.

\begin{equation}
	J'=(22\frac{2}{3}) \frac{t^4}{U^3}
\end{equation}
and 
\begin{equation}
	J'''=(3\frac{1}{3}) \frac{t^4}{U^3}
\end{equation}

\begin{equation}
	K_3=-6\frac{t^3}{U^2}e^{i\Phi_3}-30\frac{t^4}{U^3}e^{i\Phi_4}-4\frac{t^4}{U^3}
\end{equation}
\begin{equation}
	K'_3=-10\frac{t^4}{U^3}e^{i\Phi_4}-\frac{4}{3}\frac{t^4}{U^3}
\end{equation}
and
\begin{equation}
	K_4=20\frac{t^4}{U^3}
\end{equation}

For $SU(4)$, the following expression may be useful:
\begin{equation}
	P_{123}+h.c.=-\frac{1}{8}\sum_{[t_a,t_b]=0}  t_a \otimes  t_b \otimes (t_a t_b)
\end{equation}
\begin{equation}
	i(P_{123}-h.c.)=\frac{1}{16}\sum_{a,b,c}  f_{abc} t_a \otimes t_b \otimes t_c
\end{equation}
where $[t_a,t_b]=i f_{abc} t_c$. $a,b,c$ is summed over $0,1,...,15$. Note that the second term is time reversal odd.

\section{Quantum oscillation of the spinon Fermi surface}

Strictly speaking, the spinon Fermi surface is neutral and only couples to the internal $U(1)$ gauge field. However, the internal magnetic field $b=da$ can be locked to the external magnetic field\cite{motrunich2006orbital,sodemann2018quantum}:
\begin{equation}
	b=\gamma B
\end{equation}
with a factor $\gamma$ which depends on $\frac{t}{U}$. 

From Eq.~\ref{eq:mean_field_energy}, for a uniform ansatz giving spinon Fermi surfaces, the mean field energy is:
\begin{equation}
	\frac{E_M}{N}=-\tilde J\sum_{\langle ij \rangle}|\langle \hat T_{ij} \rangle|^2 -2\tilde K\sum_{i,j,k\in \bigtriangleup}|\langle \hat T_{ij} \rangle|^3 \cos(\Phi^b_3-\Phi_3^B)
	\label{eq:mean_field_energy_B}
\end{equation}
where we assume a uniform $|\langle \hat T_{ij} \rangle|$. $\Phi_3^B$ and $\Phi_3^b$ are the external and internal flux through a triangle. The $\tilde K$ term clearly favors $\Phi^b_3$ to be locked to the external flux $\Phi^B_3$. However, the diagmatic response of the spinon Fermi surface will suppress an internal magnetic flux. The competition of the above two opposite effects lead to $b=\gamma B$ with $\gamma<1$.   Because of the locking between $b$ and $B$, spin resistivity will show a Shubnikov-de Haas quantum oscillation under external magnetic field $B$.

$b$ is dynamically generated to minimize the following energy 
\begin{equation}
	E_M(B)=\chi_f |b|^2+ A |b-B|^2
\end{equation}
where $\chi_f \sim \frac{1}{m^*} \sim J$ is the Landau diamagnetism for the spinon Fermi surface and  $A\sim K$ is from the second term of Eq.~\ref{eq:mean_field_energy_B}.  The minimization leads to $\gamma=\frac{b}{B}=\frac{A}{A+\chi_f}$.  In the $K<<J$ limit, we get $\gamma \sim \frac{K}{J} \sim \frac{t}{U}$.  So deep inside the Mott insulator this effect may be small.  However, $\gamma$ is expected to become $1$ when the system is close to the Mott transition, although the effective spin model breaks down in such a regime.

\section{Details of the large-N mean field calculations}

\begin{figure}[H]
\centering
\includegraphics[width=0.85\textwidth]{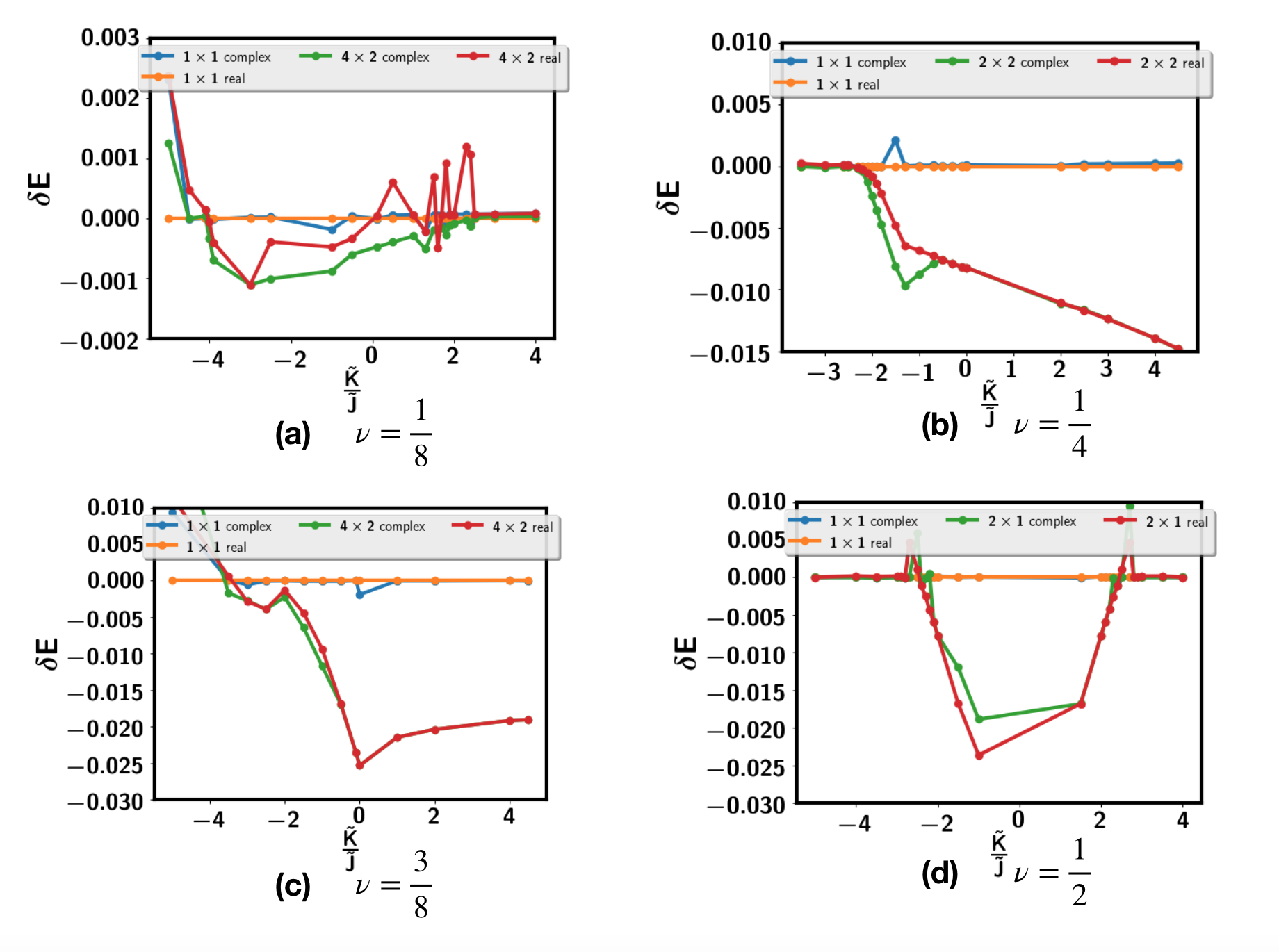}
\caption{Mean field energy for different fillings. $\delta E$ is defined relative to the energy of the ansatz with $1\times 1$ and real hopping.}
\label{fig:mean_field_energy}
\end{figure}

In the calculation, we choose different unit cell with size $m\times n$ and solve the self-consistent equations in Eq.~\ref{eq:self_consistent_equations} using the iteration method starting from a randomly chosen ansatz.  The iteration method is not guaranteed to find the global minimum. We need to start from several different initial ansatz and keep the best solution.   During every step of the iteration, $m\times n$ number of chemical potentials are solved to implement the constraint that $\frac{1}{N}\langle f^\dagger_{i;\alpha} f_{i;\alpha} \rangle=\nu$  at each site $i$.  This requires to solve $m\times n$ non linear equations, which is a very hard task.  In our calculation, we label the $m\times n$ site as $i=1,2,...,m\times n$. At every step, we get the chemical potential $\mu_i$ with the bisection method from the equation $n_i=\langle f^\dagger_i f_i \rangle=\frac{1}{\nu}$ and then moves to the next site $i+1$.  By repeating this procedure, we managed to let $|n_i-\frac{1}{\nu}|<0.0001$ for most of the cases. However, for $\nu=\frac{1}{8}$, the solution of chemical potential is not very good for the $4\times 2$ with real hopping and as a result there is noise in the mean field energy, as shown in Fig.~\ref{fig:mean_field_energy}. But the energy of the $4\times 2$ with real hopping is always higher than the CSL state even with noise, so we believe the conclusion that the CSL phase is the ground state for $\frac{\tilde K}{\tilde J} \in [-4.0,2.3]$ is robust. 

In Fig.~\ref{fig:mean_field_energy} we show the energy of different ansatz along with $\frac{\tilde K}{\tilde J}$. For each filing $\nu$, we choose different unit cells. For each unit cell configuration, we choose to restrict $t_{ij}$ to be real or allow it to be complex.  For each filling $\nu$, we tried several different unit cell, but we find one specific unit cell is the mostly favored configuration in addition to the translation invariant ansatz. For example, for $\nu=\frac{1}{8}$, the favored unit cell configuration is $4 \times 2$. In the ansatz with enlarged unit cell, usually the two lowest energy ansatz are the crystal phase and the chiral spin liquid (CSL) phase. For the crystal phase, we find the ansatz restricted to real hopping and complex hopping has the same energy, as shown for $\frac{\tilde K}{\tilde J}>-0.5$ at $\nu=\frac{3}{8}$ and $\frac{\tilde K}{\tilde J}>-0.7$ at $\nu=\frac{1}{4}$,  For both $\nu=\frac{1}{4}$ and $\nu=\frac{1}{8}$, in a region for negative $\tilde K$, the energy for the complex ansatz is lower than that with only real hopping. In this region, we find that the CSL phase has slightly lower energy than the crystal phase. For $\nu=\frac{1}{2}$, the energy of the crystal phase is lower than the CSL phase with $\frac{\tilde K}{\tilde J}\in [-2.4,2.5]$, outside of which the spinon Fermi surface state is favored. However, the CSL phase has only slightly higher energy after adding $\tilde K$. This implies that the CSL phase may be favored if we add even a small external flux $\Phi_3$ at half filling $\nu=\frac{1}{2}$ .

\begin{figure}[H]
\centering
\includegraphics[width=0.85\textwidth]{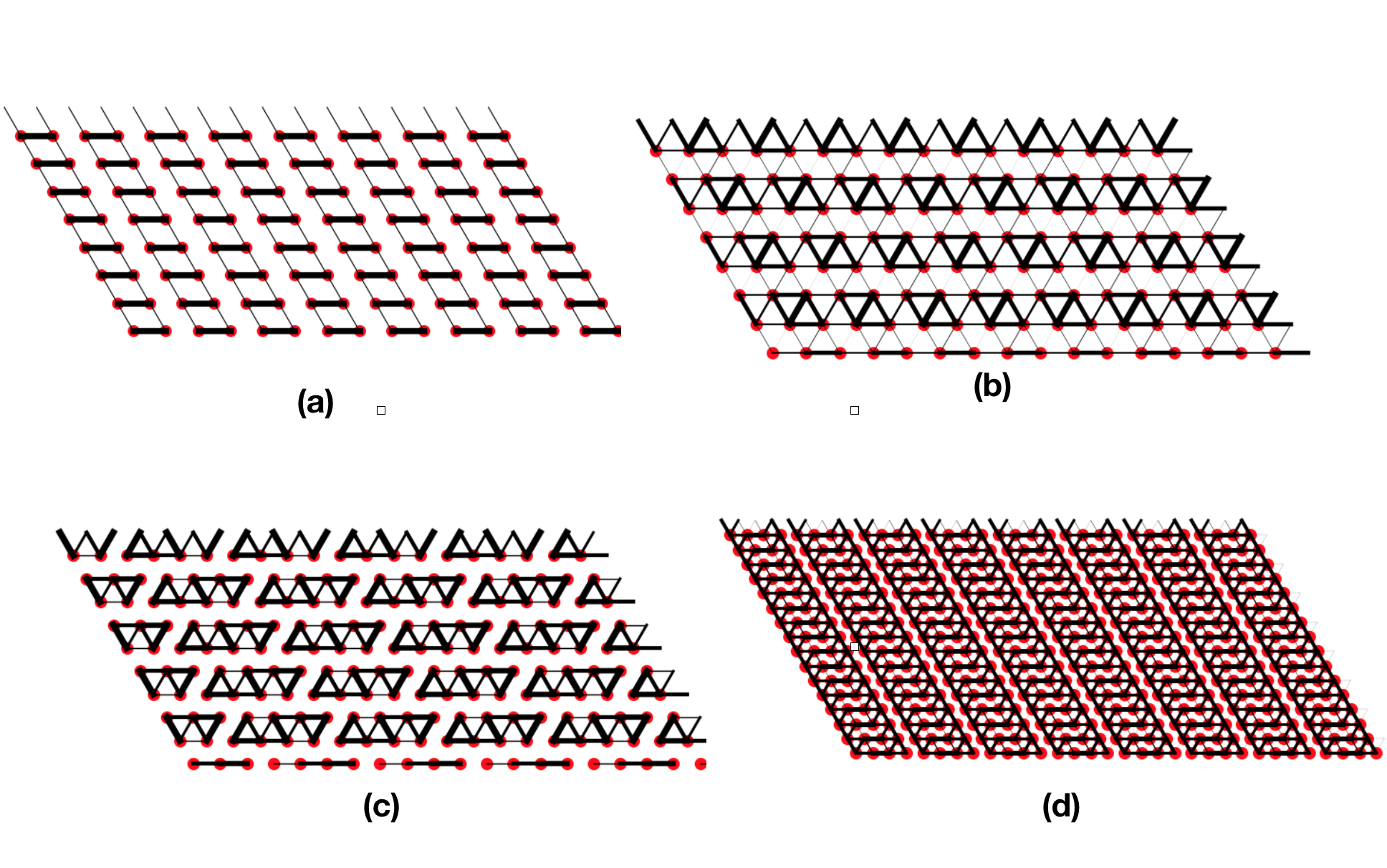}
\caption{The amplitude of the hopping $|\chi_{ij}|$ for the crystallized phase. (a) VBS order for $\nu=\frac{1}{2}$ at $\tilde K=0$. (b) Plaquette order for $\nu=\frac{1}{4}$ at $\tilde K=0$. (c) A crystallized order for $\nu=\frac{3}{8}$ at $\tilde K=0$.  (d) A "stripe" order for $\nu=\frac{3}{8}$ at $\frac{\tilde K}{\tilde J}=-2.5$, the translation symmetry breaking along $\tilde a_2$ is much weaker.}
\label{fig:crystal_plot}
\end{figure}

\begin{figure}[H]
\centering
\includegraphics[width=0.85\textwidth]{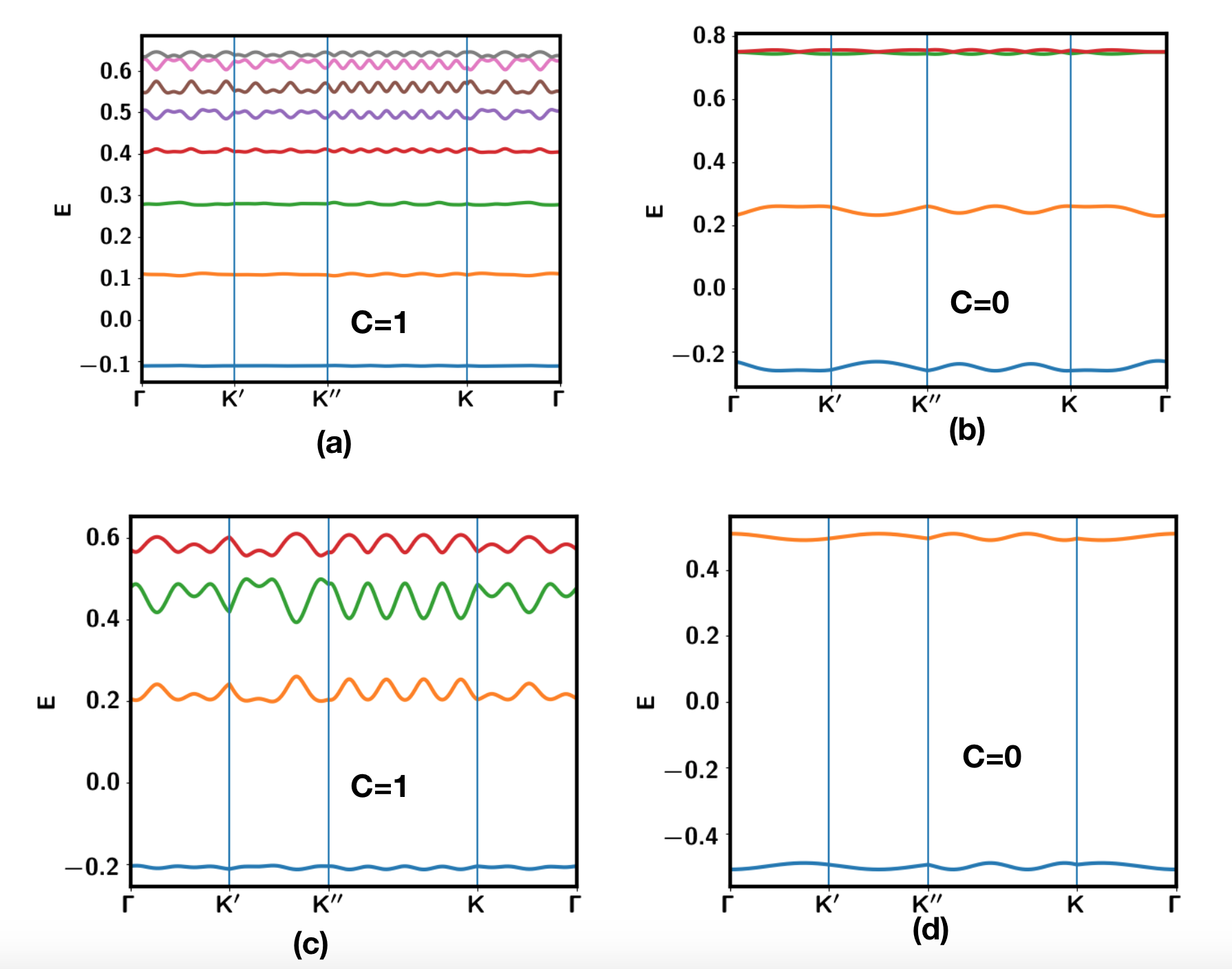}
\caption{Dispersion for crystal and CSL phases. The energy is in unit of $\tilde J$. (a) The CSL for $\nu=\frac{1}{8}$ at $\tilde K=0$; (b) The Plaquette order for $\nu=\frac{1}{4}$ at $\tilde K=0$; (c) The CSL for $\nu=\frac{1}{4}$ at $\frac{\tilde K}{\tilde J}=-1$; (d) The VBS for $\nu=\frac{1}{2}$ at $\tilde K=0$. }
\label{fig:dispersion_plot}
\end{figure}

\begin{figure}[ht]
\centering
\includegraphics[width=0.85\textwidth]{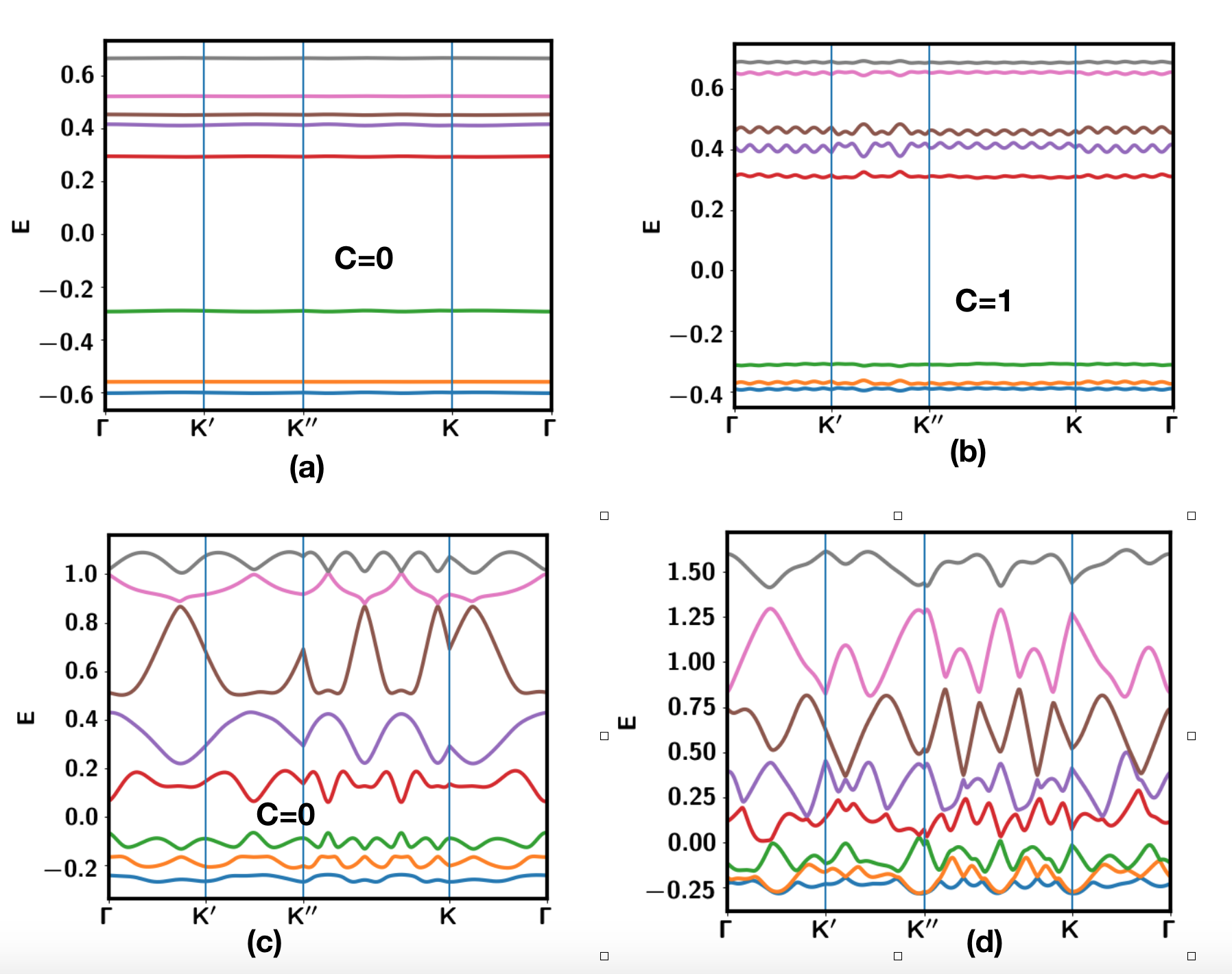}
\caption{Dispersion for $\nu=\frac{3}{8}$. The energy is in unit of $\tilde J$. (a) The crystallized phase at $\tilde K=0$; (b) The CSL phase $\frac{\tilde K}{\tilde J}=-0.5$; (c) The stripe phase at $\frac{\tilde K}{\tilde J}=-2.5$. The unit cell is still $4\times 2$, but the symmetry breaking along $\mathbf a_2$ direction is weak. (d) The stripe phase at $\frac{\tilde K}{\tilde J}=-3.5$. In this case the symmetry breaking along $\mathbf a_2$ direction is almost zero. As a result the phase is gapless along $\mathbf a_2$ direction.}
\label{fig:dispersion_plot_2}
\end{figure}

The symmetry breaking pattern for the crystal phase is shown in Fig.~\ref{fig:crystal_plot}. For the CSL phase the amplitude of the hopping is uniform. In Fig.~\ref{fig:dispersion_plot} and Fig.~\ref{fig:dispersion_plot_2} we show dispersion for the crystal and CSL phases at various fillings and different values of $\tilde K$.  One can see that the dispersion is usually quite flat and there is an obvious gap at order of $0.2-0.5 \tilde J$. For $\nu=\frac{3}{8}$, a stripe order is found for $\tilde K/\tilde J \in [-3.5,-2.0]$. In this stripe order the gap is very small or zero.

\subsection{Competing spinon Fermi surface ansatz}

For large  $|\frac{\tilde K}{\tilde J}|$, we find that spinon Fermi surface ansatz with $1\times 1$ unit cell is favored.  However, there can be several different ansatz with close energy. Their main difference is whether the $C_6$ symmetry is preserved.

\begin{figure}[ht]
\centering
\includegraphics[width=0.85\textwidth]{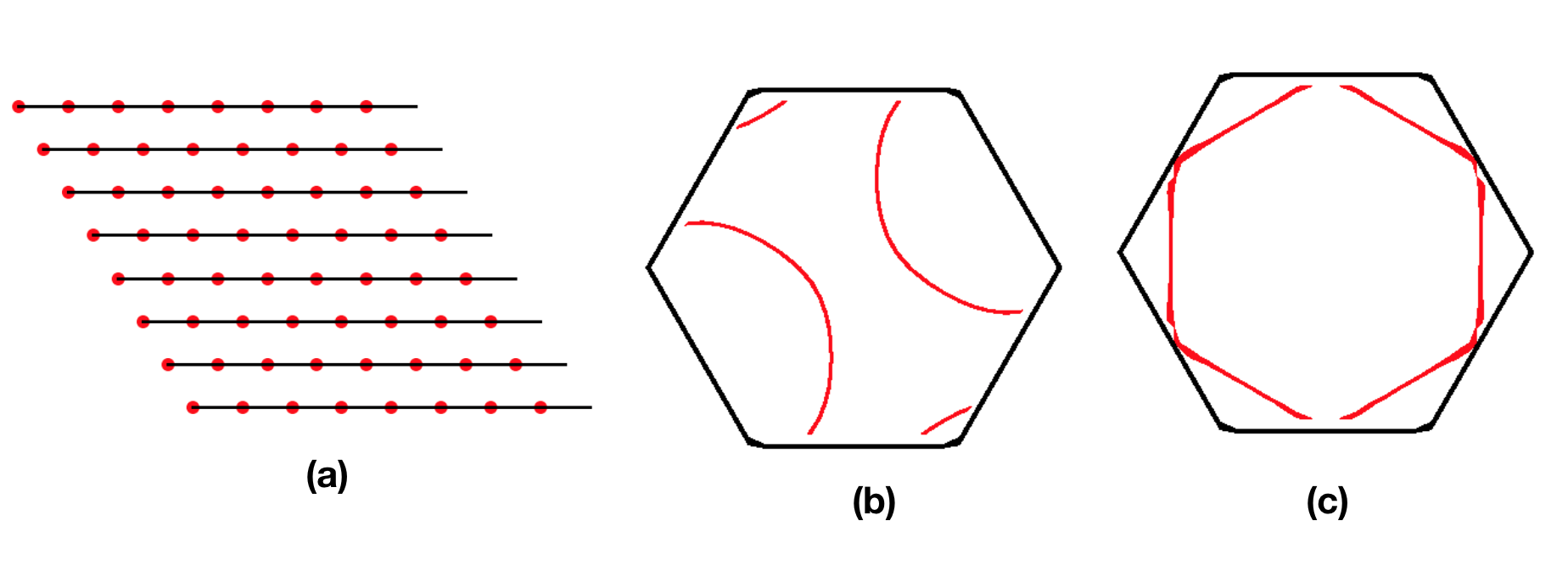}
\caption{Different competing ansatz for spinon Fermi surface with $1\times 1$ unit cell. (a) Plot of $\chi_{ij}$ for the decoupled chain phase, favored at $\frac{\tilde K}{\tilde J}<-2.3$ regime for $\nu=\frac{1}{4}$ and $\frac{\tilde K}{\tilde J}<-4.0$ for $\nu=\frac{1}{8}$. (b) Another competing spinon Fermi surface state for $\frac{\tilde K}{\tilde J}<-2.4$ at $\nu=\frac{1}{2}$; $C_6$ symmetry is broken down to $C_2$ and the center of the spinon Fermi surface is at the M point.  (c) At $\nu=\frac{1}{4}$, when we decrease $\frac{\tilde K}{\tilde J}$ to around $-5$, the  uniform ansatz wins over  the decoupled chain phase. We have $\chi_{ij}<0$, and  the  spinon Fermi surface has Van Hove singularity, the same as that for the Fermi liquid at $\nu=\frac{3}{4}$.}
\label{fig:spinon_FS_plot}
\end{figure}

For $\nu=\frac{1}{4}$ and $\nu=\frac{1}{8}$, we find decoupled chain phase is favored for negative large $\frac{\tilde K}{\tilde J}$.  But for $\nu=\frac{1}{4}$, if we further increase $|\tilde K|$, the uniform ansatz finally wins, as shown in Fig.~\ref{fig:spinon_FS_plot}(c). However, the hopping $\chi_{ij}$ is negative, implying that the Fermi surface is the same as that for $\nu=\frac{3}{4}$ if we assume $t>0$ in the Hubbard model. There is a van Hove singularity and the state should be unstable.  The weak Mott regime of $\nu=\frac{3}{4}$ is likely to be a spin density wave insulator.  For $\nu=\frac{1}{2}$, we also find a competing spinon Fermi surface phase shown in Fig.~\ref{fig:spinon_FS_plot}(b).  At each site $i$, there are three hoppings along three directions.  In this nematic phase with only $C_2$ symmetry, all of three hoppings have the same magnitude. However, two of them are positive while the third one is negative.

In the region where spinon Fermi surface state is favored, $t/U$ should already be large and our calculation at the order $O(\frac{t^3}{U^2})$ may not be sufficient to find the ground state. We leave a better analysis to future.

\end{document}